\documentclass[onecolumn,amsmath,amssymb,aps,prd,showpacs,showkeys,nofootinb ib,floatfix]{revtex4}

\usepackage[dvips]{graphicx}
\usepackage[dvips]{color}

\begin{document}

\newcommand{\beq}{\begin{eqnarray}}
\newcommand{\eeq}{\end{eqnarray}}
\newcommand{\be}{\begin{equation}}
\newcommand{\ee}{\end{equation}}

\def\la{\mathrel{\mathpalette\fun <}}
\def\ga{\mathrel{\mathpalette\fun >}}
\def\fun#1#2{\lower3.6pt\vbox{\baselineskip0pt\lineskip.9pt
\ialign{$\mathsurround=0pt#1\hfil ##\hfil$\crcr#2\crcr\sim\crcr}}}

\newcommand{\veX}{\mbox{\boldmath${\rm X}$}}
\newcommand{{\SD}}{\rm SD}
\newcommand{{\Lc}}{\mathcal{L}}
\newcommand{{\Mc}}{\mathcal{M}}
\newcommand{\pp}{\prime\prime}
\newcommand{\veY}{\mbox{\boldmath${\rm Y}$}}
\newcommand{\vex}{\mbox{\boldmath${\rm x}$}}
\newcommand{\vey}{\mbox{\boldmath${\rm y}$}}
\newcommand{\vef}{\mbox{\boldmath${\rm f}$}}
\newcommand{\ver}{\mbox{\boldmath${\rm r}$}}
\newcommand{\vesig}{\mbox{\boldmath${\rm \sigma}$}}
\newcommand{\vedelta}{\mbox{\boldmath${\rm \delta}$}}
\newcommand{\veP}{\mbox{\boldmath${\rm P}$}}
\newcommand{\vep}{\mbox{\boldmath${\rm p}$}}
\newcommand{\veq}{\mbox{\boldmath${\rm q}$}}
\newcommand{\vez}{\mbox{\boldmath${\rm z}$}}
\newcommand{\veS}{\mbox{\boldmath${\rm S}$}}
\newcommand{\veL}{\mbox{\boldmath${\rm L}$}}
\newcommand{\veK}{\mbox{\boldmath${\rm K}$}}
\newcommand{\veR}{\mbox{\boldmath${\rm R}$}}
\newcommand{\ves}{\mbox{\boldmath${\rm s}$}}
\newcommand{\vek}{\mbox{\boldmath${\rm k}$}}
\newcommand{\ven}{\mbox{\boldmath${\rm n}$}}
\newcommand{\veu}{\mbox{\boldmath${\rm u}$}}
\newcommand{\vev}{\mbox{\boldmath${\rm v}$}}
\newcommand{\veh}{\mbox{\boldmath${\rm h}$}}
\newcommand{\verho}{\mbox{\boldmath${\rm \rho}$}}
\newcommand{\vexi}{\mbox{\boldmath${\rm \xi}$}}
\newcommand{\vepi}{\mbox{\boldmath${\rm \pi}$}}
\newcommand{\veta}{\mbox{\boldmath${\rm \eta}$}}
\newcommand{\veB}{\mbox{\boldmath${\rm B}$}}
\newcommand{\veH}{\mbox{\boldmath${\rm H}$}}
\newcommand{\veE}{\mbox{\boldmath${\rm E}$}}
\newcommand{\veJ}{\mbox{\boldmath${\rm J}$}}
\newcommand{\veal}{\mbox{\boldmath${\rm \alpha}$}}
\newcommand{\vegam}{\mbox{\boldmath${\rm \gamma}$}}
\newcommand{\vepar}{\mbox{\boldmath${\rm \partial}$}}
\newcommand{\llan}{\langle\langle}
\newcommand{\rran}{\rangle\rangle}
\newcommand{\lan}{\langle}
\newcommand{\ran}{\rangle}

\author{Yu.A.Simonov}
\email{simonov@itep.ru}

\affiliation{ITEP, Moscow, Russia}

\title{Pattern of fermion masses from  high-scale evolution}


\begin{abstract}
Dynamical  equations for fermion masses are derived using high scale universal
mass generation  and consequent mass evolution due to $SU(3), SU(2)$,  and
$U(1)$ gauge interaction. Assuming    mass generation at the GUT scale
$M=10^{14}$ GeV, one obtains hierarchy and  a large spread in fermion masses
with roughly correct values of $m_\nu, m_\tau, m_t, m_b$ in the third
generation. The smallness of neutrino mass, $\nu_3\sim 10^{-12} m_{t}$,
naturally arises in the solution.

\end{abstract}


\maketitle



{\bf 1.} The problem of fermion masses is being studied for many years (see
\cite{1} for reviews and references). The most striking points are  large
spread and hierarchy of masses both in  vertical  (inside one generation) and
horizontal directions(i.e. from one generation to another), and  also the
extreme smallness of neutrino mass.

In the Standard Model (SM) scenario fermion mass generation is  related to the
Yukawa Higgs constants. When one tries to understand  fundamental dynamics
behind Higgs field and express  all  effects in terms of  fields at high scale
and known gauge fields, one realizes that visible masses at our scale ($\sim 1$
GeV) are due  to several sources.

 First of all,   resulting
fermion masses are to be created in the  original chiral symmetry breaking
(CSB) process (possibly at high scale), and    then  they are evolved by all
known gauge interactions,  and finally (or originally) mixed and shifted by
general CKM mechanism.

In  recent  publications \cite{2,3,4} the author  has argued, that the original
mass generation process can be associated with   CSB due to topological charges
in  the electroweak (EW) vacuum. This process  is similar to CSB  in the
instanton gas, which was studied in different approaches in QCD  \cite{5,6}; in
what follows we shall use the
  formalism of \cite{7,8}\footnote{Strictly  speaking, both CSB and mass generation mechanisms are not  derived in \cite{5,6,7,8}
  in  a gauge invariant formalism, sice the instatnton gas with net zero
  topcharge is  introduced in a certain gauge. A gauge invariant mass
  generation mechanism (outside of elementary Higgs  model and confining phase of QCD) is not known to the
  author}. General setting of the problem is given in
\cite{4} in the framework of the Pati-Salam  $G(2,2,4)$ group \cite{9}, but for
  present paper the details of $SO(10)$ group, which is splitted down to
$G(2,2,4)$, are not important, and one can use the $SU(2)$ instanton as the
basic element of $SO(2n)$ or $SU(n)$ group \cite{10}. The interesting feature
of the $SU(2)$ instanton (or any local topcharge) is that it produces  integral
equation with the  kernel of the same structure as  in the fermion self-energy
equations   \cite{11,12}, (see \cite{13}  for review and  earlier references)
\footnote{The main emphasis of \cite{11,12,13} is on the possible new type of
evolution, given by a linearized equation, while for the present analysis the
first solution and  the nonlinear regime  are relevant.},
 with the
natural cutoff at the mass $M\approx 1/\rho$, where $\rho$ is the topcharge
radius (see appendix 2 of \cite{4}). The basic point  is that the resulting
nonlinear integral equation allows to express   fermion masses through the
high-scale cutoff mass $M$.

\vspace{1cm}

{\bf 2. }  We start with   general equation describing the process of CSB and
mass generation at some high scale $M$ and the consequent mass evolution. For
the Euclidean momentum-dependent fermion mass $\mu_i (p) $, one can write

 \be
\mu_i (p) = \int^M \frac{b_{i} (q) \mu_i (p_1) d^4 p_1}{q^2(p^2_1 + \mu^2_i
(p_1))}, ~~ q\equiv  p-p_1.\label{1}\ee Here $i$ refers to fermions  within the
highest generation  $i=(\nu,\tau, t,b)$. Taking into account, that the mass of
each fermion is generated with  the interaction constant $b_i^{(0)}(q)$ and is
subject to evolution due to gauge fields of group
 $SU(3)_c \times U(1)_{em}$, one can write

\be  b_i (q) = b_i^{(0)}  +\sum_{n=1, 3}
\nu_n^{(i)}\frac{\alpha_n(q)}{\pi}\label{3}\ee where $\nu_n^{(i)} $ is the
weight of charge  $n$ for fermion $i$. Here we shall consider only the third
generation to avoid  confinement complications at low scale,  hence
$\nu_1^{(i)} = (0,1, \frac49, \frac19)$  and $\nu_3^{(i)} = (0,0,1,1)$ for
$(\nu, \tau, t, b)$, while $\nu_2^{(i)}$ is calculated from the $Z_0$ exchanges
and will be neglected in the first approximation. The constant $b^{(0)}_i$ is
proportional to topcharge density and  depends on  $i$ in the $SU(2)$ broken
vacuum \cite{4}.

For $\alpha_n (q^2)$ one can use the one-loop evolution,

\be \frac{\alpha_n(q^2)}{\pi} =\frac{c_n}{1+ \omega_n \ln q^2/M^2}, ~~ \omega_n
= \frac{\beta_0^{(n)}}{4\pi} \alpha_n (M^2),\label{4}\ee where $c_n
=\frac{1}{\pi} \alpha_n (M^2), ~~  \beta_0^{(n)} = \frac{11 n - 2 n_f}{3}, ~~
n=3,$ and  $\beta_0^{(1)} =- \frac{4n_f}{3}$. For $\alpha_3(q)$ we implicitly
introduce IR freezing at small $q^2$, (see \cite{14} for review and
references),  which   contributes less than 10\%  for the third generation.

Integrating in (\ref{1}) over angles and introducing $s\equiv p^2/M^2$ and
$\kappa (s) \equiv \mu(p^2/ M^2) / M$, one obtains integral equation

\be \kappa_i(s)  =  \int_0^1 \frac{b_i (\sigma) \kappa_i (s_1) s_1
ds_1}{\sigma(s_1+\kappa_i^2(s_1))}, ~~ \sigma=\max (s,s_1),\label{5}\ee where
\be b_i (s) = b_i^{(0)} + \sum^3_{n=1} \nu_n^{(i)} \frac{\alpha_n(s)}{\pi}, ~~
i= \nu,\tau, t, b\label{6} \ee

differential equation \cite{11,12,13}, (up to small terms $O(\omega_n)$). \be
s^2 \kappa_i^{''}(s) + 2s \kappa_i' (s)+ \frac{sb(s) \kappa_i(s)}{s+ \kappa^2_i
(t)} =0.\label{7}\ee

For a constant $b_i\equiv b$ and for $s\gg \kappa^2_i$ one finds two solutions
$\kappa(s) = s^\delta,~~ \delta_1 =-\frac12 + \sqrt{\frac14 -b} \approx -b$,
$~\delta_2 \approx -1 +b$. In what follows only the first solution will be
appropriate  in the integral equation (\ref{5}) for small $s$. For $s\la
\kappa^2 (0)$ one has a solution of nonlinear  equation (\ref{7}) \be \kappa_i
(s) \simeq \sqrt{\kappa^2_i(0)- b_is}.\label{8}\ee Now taking into account
 the evolution of $b_i(s)$, given by (\ref{6}), (\ref{4}), the solution of the  linearized
Eq. (\ref{7}) acquires the form \be \kappa_i(s) = \kappa_i(s_0) \left(
\frac{s}{s_0} \right)^{ - b_i(0) }\prod_{n=1,2,3} \left( \frac{1+\omega_n  ln
s}{1+\omega_n \ln s_0}\right)^{-a_n^{(i)}}, ~~s\geq s_0\label{9}\ee with
$a^{(i)}_n = \nu_n^{(i)}\frac{c_n}{\omega_n}$.  We are now matching two
solutions  and their derivatives, Eq. (\ref{8}) for $s\leq s_0$, and Eq.
(\ref{9}) for $s\geq s_0$, which yields $s_0 \approx 2 (\kappa_0^{(i)})^2,  ~~
\kappa_0^{(i)} \equiv \kappa_i (0)$. To express $\kappa_0^{(i)}$ through $M$
(which is the only mass parameter of our problem), one can insert the matched
solution into Eq. (\ref{5}) at $s=s_0$, which yields \be b_i^{(0)} \ln s_0 +
\sum_n a_n^{(i)}\ln (1+\omega_n \ln s_0) = f_i ,\label{10}\ee where $f_i \equiv
\ln (b_i (s_0) 0.45), ~ s_0 = 2 (\kappa_i ((0))^2$.

In the simplest approximation, when $\ln (1+\omega_n \ln s_0) \approx \omega_n
\ln s_0$ the solution, for fermion mass $\mu_i^{(0)}\equiv \mu_i (p^2= 2
(\mu_i^{(0)})^2)$ is \be 2\left( \frac{\mu^{(0)}}{M}\right)^2 =\exp \left(
\frac{f_i}{b_i^{(0)}+ \sum_{n=1}^3 \nu_n^{(i)} \frac{\alpha_i
(M)}{\pi}}\right).\label{11}\ee In general case one has instead \be 2\left(
\frac{\mu^{(0)}}{M}\right)^2 =\exp \left( \frac{f_i- \sum_{n=1}^3 a_n^{(i)} \ln
(1+ \omega_n \ln s_0)}{b_i^{(0)}}\right).\label{12}\ee

Both forms, (\ref{11}) and (\ref{12}) demonstrate an extreme sensitivity of the
resulting fermion mass to the parameter $b_i^{(0)}$. The coefficients
$\nu_n^{(i)}$, which define both the spread and the hierarchy of fermion masses
$\mu_i^{(0)}$,  are  discussed  below.

\vspace{1cm}

{\bf 3.} To make numerical predictions, we  start with the mass  $M=10^{14}$
GeV, which is around the point  where all  constants  $\alpha_n(M)$ are nearly
intersecting, as it happens in $SU(5)$ and $SO(10)$ groups  \cite{15}: so we
take $\alpha_n(M)=1/43, ~~ M=10^{14}$ GeV, and $\alpha_1(M)\simeq 0.01$.
Variations of $M$ in the range $10^{14}\div 10^{16}$ GeV with unequal $
\alpha_n(M)$ do not change results qualitatively, if  the appropriate change in
$b^{(0)}$ is made. For simplicity we also neglect  contribution of $\alpha_2$,
which can be compensated by a  small change $0(10^{-4})$ in $b^{(0)}$. As a
first approximation we consider an unbroken $SU(2)$ at high scale with a common
$b^{(0)}$. We have chosen $b^{(0)}$ in the interval [0.03; 0.05]; results for
the  masses $\mu_i$   are shown in Table 1  and  compared with experimental
values for the third generation. Note, that due to very high sensitivity of Eq.
(\ref{12}) to   entering numbers, the accuracy of results in Table  1 is low
and the entries are rather indicative of orders of magnitude.

The first thing is to check, whether (\ref{11}), (\ref{12}) predict   correct
hierarchy within the generation $(\mu_t >\mu_d > \mu_b> \mu_\tau)$, when one
keeps $b^{(0)}$ constant, i.e. whether the hierarchy  is due to  $SU(3)_c\times
 U(1)_{em}$ evolution. Looking at   Table
1, one indeed can see  that the mass  hierarchy is kept correct for all values
of $b^{(0)}$. From  Eq. (\ref{11}), one can understand  why the hierarchy is
natural in our approach:what enters  in the denominator of the exponent is  the
weighted sum of all charges squared, which is  the smallest for $\nu$ and
increasing for $\tau, b $, and finally  is the largest for $t$. Now, since
$f_i$ is negative, this hierarchy of denominators is zoomed up in the resulting
mass values. Moreover, this enhancement is so high, that  the $\sim 30\%$
change of the denominator value may produce ten orders of magnitude in mass
values (e.g. for $b^{(0)}\approx 0.035)$. This explains why for $b^{(0)}=
0.035$ one obtains $\mu_{\nu_3} \sim 10^{-12}$ $\mu_{t}$.  This ``zoom effect''
can explain small  Dirac neutrino masses without any extra mechanisms.

Thus two  main properties of fermion spectra: the hierarchy and the large mass
spread are qualitatively reproduced by the simplest variant of our model with a
common $b^{(0)}$.

However, the absolute values and mass ratios are in many cases  far from
experiment. Especially the ratios $\mu_\nu/ \mu_\tau$ and $\mu_\tau/\mu_b$ are
nine orders of magnitude off the experimental values. To improve agreement we
shall take into account the $SU(2)$ splitting, discovered in \cite{4}, which
splits $ b_t^{(0)}$  from $b_b^{(0)}$, and $b_\nu^{(0)}$ from $b_\tau^{(0)}$.

Results of this analysis with the central value $b^{(0)}=0.045$ are positive in
the sense, that indeed all four masses of the third generation are well
reproduced, when $b_i^{(0)} = b^{(0)} + \Delta b_i$, where $\Delta b_i\simeq (-
0.009;  + 0.01; + 0.004; -0.004)$ for $(\nu, \tau, t,b)$. It is remarkable,
that the structure $ \Delta b_i \cong (- 2\varepsilon; + 2 \varepsilon;
\varepsilon; -\varepsilon), \varepsilon \ll b^{(0)}$ may be indicative of a
certain $SU(2)$ symmetry group violation. A similar analysis for $M=10^{10}$
GeV allows to reproduce experimental masses with $b_1= b^{(0)}+\Delta b_i,
b^{(0)}= 0.07, \Delta b_i =(-0.028; 0; 0; -0.007)$.

 Till now we have considered only the heaviest, third generation. Lower
 generations  can be also   included, they correspond to the splitted
 quadruplets centered around $b^{(0)}=0.036$ and $b^{(0)}=0.03$ for the  second
 and first generations respectively. This calculation, however, does not explain
 the generation mechanism, but rather readdresses  it  to the high scale
 dynamics, where $\mu_i$ and $b_{i}$ in Eq. (\ref{1}) must become matrices in
 generation indices, which naturally  provide both masses and mixing
 coefficients. This point is now under investigation.

 It is also interesting, that the same equation (\ref{1}) would result if the
 high-scale mass generation is given without top charge by mechanism of gauge
 field exchanges, in which case $b_i^{(0)} $ is  the squared charge
 $\nu^{(i)}_{n'} \frac{\alpha_{n'}}{\pi}$ for  this new interaction, and  $M$ is
 the scale, where new interaction splits off  from the original GUT.

 Summarizing, we have derived dynamical equation for fermion masses, where the
 $SU(3) \times U(1) $ evolution  and the (10-20)\% spread in  $b_i^{(0)}$  naturally explains the hierarchy and
the  spread of masses within one generation, and especially the smallness of
 neutrino masses  through     the model parameter --  the topological
 mass generation
constant $b^{(0)}$. With a small variation of this parameter the exact masses
of the third generation are reproduced. Of course, our discussion above is
oversimplified. We have ignored processes like  breaking  of $SU(4)_{lc}$  and
neglected   $SU(2)_R$  and  the $U(1)_Y$ contribution, stressing only
qualitative mechanism. The explicit type of gauge interaction yielding constant
$b_i^{(0)}$, was not specified in the paper , and it is an open question,
whether it would pass precise EW tests.  The author is grateful for financial
support to the grant RFBR no. 09-02-00620a.

\begin{table}
\caption{Fermion masses (in GeV/$c^2$) of the third generation  for different
the topological constant $b^{(0)}$ in comparison with experimental values }
\begin{center}

\begin{tabular}{|c|c|c|c|c|c|}
\hline $b^{(0)}$   & 0.03&0.035&0.0442&0.05& experiment\\\hline &&&&&\\

 $\mu_{\nu_3}$&
$0.6\cdot 10^{-17}$& $0.16\cdot 10^{-12}$& $0.5 \cdot
10^{-5}$& $3.1\cdot 10^{-3} $& $5 \cdot 10^{-11}$\\

$\mu_{\tau}$&  $0.17\cdot 10^{-14}$& $2\cdot 10^{-10}$& $1.1\cdot 10^{-4}$& $0.034$& $1.78$\\

$\mu_t$& 0.009& 0.5&24&$214$ &174\\
$\mu_b$ &0.0025& 0.18&12& 110& 4.2\\
\hline

\end{tabular}

\end{center}

\end{table}

\end{document}